\documentstyle[12pt,epsfig]{article}
 \newcommand{\be}{\begin{eqnarray}}
 \newcommand{\ee}{\end{eqnarray}}
 \newcommand{\beq}{\begin{equation}}
 \newcommand{\eeq}{\end{equation}}
 \newcommand{\ba}{\begin{array}{1}}
 \newcommand{\ea}{\end{array}}
 \newcommand{\bb}{\begin{thebibliography}}
 \newcommand{\eb}{\end{thebibliography}}

 \newcommand{\abstitle}[1]{{\small {\bf #1}}}
 \newcommand{\absauthor}[1]{{\small {\bf #1}}}
 \newcommand{\address}[1]{{\it #1}}
 
\begin{document}
 \begin{center}
 \abstitle{Forward heavy flavour production in $pp$ collisions at LHC
and intrinsic quark components in proton}\\
 \vspace{0.6cm}
 \absauthor{G. I. Lykasov$^1$, V. A. Bednyakov$^1$, A. F. Pikelner$^1$ 
 and N. I. Zimin$^{1,2}$} 
\\ [0.6cm]
 \address{$^1$ Joint Institute for Nuclear Research -
  Dubna 141980, Moscow region, Russia\\
$^2$CERN, 1211, Geneva 23, Switzerland
}
 \end{center}
 \vspace{0.1cm}
%Version of April 27, 2009\\
 \vspace{0.2cm} 
\begin{center}
{\bf Abstract}
\end{center}
 \vspace{0.1cm}
The LHC data on the forward heavy flavour hadron production can be a new unique source for 
estimation of intrinsic charm and bottom contributions to the proton. For example, we analyze 
the forward heavy baryon production, namely $\Lambda_b$-baryon, within the soft QCD quark gluon
string model and present the predictions for observables which could be measured at the LHC.   
We also present some predictions for the $D$-meson production in $pp$ collisions made within the
perturbative QCD including the intrinsic charm in the proton that can be verified at the LHC.
\section{Introduction}
\label{1}
      There are successful phenomenological approaches to description of  
      the soft hadron-nucleon,
      hadron-nucleus and nucleus-nucleus interactions at high energies 
      based on the Regge theory and the $1/N$ expansion in QCD.
      For example, they are the quark-gluon string model (QGSM) 
\cite{Kaidalov:1982xg}
%{Lykasov:2009zd}
      and the dual parton model (DPM) 
\cite{Capella:1992yb}. 
      The main components of these models 
      are the parton distributions in hadrons (PDF) and the 
      fragmentation functions (FF), which describe fragmentation of quarks 
      to hadrons. The PDF and FF are expressed in terms of intercepts 
      of the Regge trajectories $\alpha_R(0)$.
 
      In these models 
      the largest uncertainty in the calculations of the %cross sections for the 
      yields of heavy flavours is %mainly 
      due to the absence of any reliable information on the 
      transfer momentum $t$ dependence
      of the Regge trajectories of heavy quarkonia ($Q{\bar Q}$).
      If teh $Q{\bar Q}$ trajectories as assumed to be linear,
%      Assuming as usual linearity of the $Q{\bar Q}$-trajectories, 
      the intercepts turn out to be low, for example, $\alpha_\psi(0)$
      is around -2
%$=-2.2$,
      and $\alpha_\Upsilon(0)$ is around -8. 
%$=-8, -16$. 
      As a result, the yield is very uncertain and  
%%%110211-> contribution of the peripheral mechanism 
       decreases very rapidly with increasing quark mass. 
%       Accordingly, 
       Furthermore, any %the finding 
       knowledge of the $t$-dependence for
       $\alpha_{(Q{\bar Q})}(t)$ in the region 
       $0\leq t\leq M^2_{(Q{\bar Q})}$ and estimations of their 
       intercepts become especially important for quantitative predictions.

       On the other side, to reduce the above-mentioned uncertainties, 
       fix the Regge trajectories of the bottom $b{\bar b}$-mesons 
       and get information about quark and diquark fragmentation functions 
       %of all quarks and diquarks 
       into heavy baryons  
       one should compare relevant QGSM predictions 
       with the LHC experimental data.
% for example 
%       on the $\Lambda_b$ production at small $p_t$ in the $pp$-collisions. 
       Therefore the first goal of this paper is to present and discuss 
       some predictions, for example, for the production of the beauty $\Lambda_b$-baryon 
       and charmed $D$-meson in $pp$ collisions. 
       The second goal of this paper is to discuss the possibility of observing  
        the so-called {\it intrinsic} quark components in $pp$ collisions 
%the proton
       at the LHC energies \cite{Pumplin:2005yf}-\cite{Thomas:1997}
%\cite{Pumplin:2005yf,Pumplin:2007wg,Nadolsky:2008zw,Litvine:1999sv,Vogt:2000,Navarra:1996,Thomas:1997}.  
       The idea of the intrinsic charm existence in the 
       proton was first put forward thirty years ago by S.Brodsky with coauthors 
\cite{ Brodsky:1980pb,Brodsky:1981}. 
\begin{figure}[h]
%\centerline{\includegraphics[width=1.0\textwidth]{nucleon.eps}}
\centerline{\includegraphics[width=0.30\textwidth]{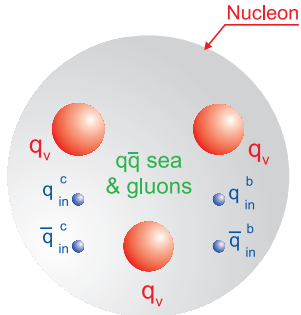}}
\caption{Schematic presentation of a nucleon consisting of three valence quarks
         q$_{\rm v}$, quark-antiquark q${\bar {\rm q}}$ and gluon sea, 
	 and pairs of the intrinsic charm 
	 (q$_{\rm in}^{\rm c}{\bar {\rm q}_{\rm in}^{\rm c}}$) 
	 and intrinsic bottom quarks 
	 (q$_{\rm in}^{\rm b}{\bar {\rm q}_{\rm in}^{\rm b}}$). %($q_{in}^b{\bar q}_{in}^b$)
}
\label{Fig_1}
\end{figure}
      They assumed the 5-quark state $uudc{\bar c}$ in the proton 
(Fig.~\ref{Fig_1}).
       Later some other models were developed. 
       One of them assumes the quasi-two-body state 
       ${\bar D}^0(u{\bar c}){\bar\Lambda}_c^+(udc)$ in the proton, 
       see for example
\cite{Pumplin:2005yf} and references therein.   
%       As is shown in \cite{Pumplin:2005yf}, 
       In \cite{Pumplin:2005yf}-\cite{Nadolsky:2008zw} 
%\cite{Pumplin:2005yf,Pumplin:2007wg,Nadolsky:2008zw}  
       a probability to find the intrinsic charm in the proton was assumed to be
       from 1 to 3.5 percent. 
%\cite{Pumplin:2005yf}, 
       The probability of the intrinsic bottom in the proton 
       is suppressed by a factor
       $m^2_c/m^2_b\simeq 0.1$ 
\cite{Polyakov:1998rb}, where $m_c$ and $m_b$ are the masses of the charmed and bottom quarks. 
       Nevertheless,
       it was also shown 
       that the intrinsic charm in the proton can result in a sizable contribution 
       to the forward charmed meson production
\cite{Goncalves:2008sw}.

       When the distribution of the intrinsic charm or bottom in the 
       proton is similar to the valence quark distribution,  
       then the production of the charmed (bottom) mesons or charmed (bottom) 
       baryons in the fragmentation region 
       is similar to the production of pions or nucleons. 

       However, the amount of this production yield depends on the probability 
       to find the intrinsic charm or bottom in the proton, but this 
       amount looks too small.    
       In this paper we continue our study of the forward heavy flavour production in $pp$ collisions at 
       LHC energies published in \cite{Bednyakov:2010yz,Lykasov:2009zd} and present some estimations 
       for the contribution
       of the intrinsic charm and bottom to the inclusive spectra of charmed and beauty baryons and mesons.
      In the next section the general formalism based on the QGSM for the 
      forward hadron production in $pp$ collisions at high energies is presented briefly. 
      In the first part of the section {\bf Results and discussion}, we present the predictions 
      for the $\Lambda_c$ and $\Lambda_b$ production in $pp$ collisions
      at the LHC obtained within the QGSM without inclusion of the intrinsic charm and bottom in the proton. 
      Then, in the second part of this section some estimations of the intrinsic charm and bottom contributions 
      to the forward production of heavy mesons and baryons in $pp$ collisions are discussed. 
      Finally, we give some recommendation for the forward LHC experiments to find the information on 
      the quark intrinsic components in the proton.

%}      The estimation of this effect in given in the paper.
 
\smallskip

%%%%%%%%%%%%%%%%%%%%%%%%%%%%%%%%%%%%%%%%%
\section{General formalism of the QGSM}
\begin{figure}[h]%\vspace{6pt}
\centerline{\includegraphics[scale=0.8]{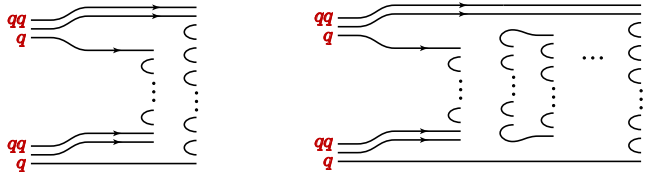}}
\caption{The one-cylinder graph (left) and the multicylinder 
graph (right) for the inclusive $p p\rightarrow h X$ process.}
\label{Fig.2}
\end{figure}

%%%%%%%%%%%%%%%%%%%%%%%%%%%%%%%%%%%%%%%%%%%%%%%%%%%%
      The general form for the invariant inclusive hadron spectrum
      within the QGSM 
\cite{Lykasov:2009zd,Capella:1992yb} is 
\begin{eqnarray}
E\frac{d\sigma}{d^3{\bf p}}\equiv
\frac{2E^*}{\pi\sqrt{s}}\frac{d\sigma}{d x d p_t^2}=
\sum_{n=1}^\infty \sigma_n(s)\,\phi_n(x,p_t)~, 
\label{def:invsp}
\end{eqnarray}
       where $E,{\bf p}$ are the energy and the three-momentum of the
       produced hadron $h$ in the laboratory system (l.s.); 
       $E^*,s$ are the energy of $h$ and the square of the initial energy in the
       c.m.s of the $pp$-system; 
       $x,p_t$ are the Feynman variable and the transverse
       momentum of $h$; 
       $\sigma_n$ is the cross section for production of
       the $n$-Pomeron chain (or $2n$ quark-antiquark strings) decaying
       into hadrons, which are calculated within the quasi-``eikonal approximation''
\cite{TerMartirosyan:1973yn}; 
        $n=1$ corresponds to the left graph in 
Fig.~\ref{Fig.2}, 
        and $n>1$ corresponds to the right graph in 
Fig.~\ref{Fig.2}. 

        The function $\phi_n(x,p_t)$ is the convolution
	of the quark (diquark) distributions and the FF 
\cite{Lykasov:2009zd,Capella:1992yb,Kaidalov:1985jg}:
\be 
\label{Eq_2}
\phi^{p p}_n(x)=F_{qq}^{(n)}(x_+)F_{q_v}^{(n)}(x_{-})+
F_{q_v}^{(n)}(x_+)F_{qq}^{(n)}(x_-)+
\\
\nonumber
2(n-1)F_{q_s}^{(n)}(x_+)F_{{\bar q}_s}^{(n)}(x_-)~,
\ee
where
$x_{\pm}=\frac{1}{2}(\sqrt{x^2+x_t^2}\pm x)$~, 
and
\be
F_\tau^{(n)}(x_\pm)=\int_{x_\pm}^1 dx_1 f_\tau^{(n)}(x_1)G_{\tau\rightarrow h}
\left(\frac{x_\pm}{x_1}\right)~.
\label{def:Ftaux}
\ee
        Here $\tau$ means the flavour of the valence (or sea) quark or diquark, 
	$f_\tau^{(n)}(x_1)$ is the quark distribution function depending on the 
	longitudinal momentum fraction $x_1$  
	in the $n$-Pomeron chain; 
	$G_{\tau\rightarrow h}(z)= z D_{\tau\rightarrow h}(z)$, 
	$ D_{\tau\rightarrow h}(z)$ is the FF of a quark (antiquark) or 
	diquark of flavour $\tau$ into a hadron $h$ (charmed or bottom hadron in our case),        
        where $z=x_\pm/x_1$. 
        The PDFs and FFs used by the calculations of Eqs.(\ref{Eq_2},\ref{def:Ftaux}) 
        are expressed
        in terms of the Regge trajectories, their intercepts $\alpha_R(0)$ and slopes 
        $\alpha^\prime_R(0)$
        All the details of the calculations
%of Eq.(\ref{def:invsp}) and the interaction function $\phi_n(x,p_t)$ 
	can be found in 
\cite{Lykasov:2009zd,Bednyakov:2010yz,Bednyakov:2011hj}.

%%%%%%%%%%%%%%%%%%%%%%%%%%%%%%%%%%%%%%%%%%%%%%%%%%%%%%%%%%%%%
\section{Results and discussion}
%%%%%%%%%%%%%%%%%%%%%%%%%%%%%%%%%%%%%%%%%%%%%%%%%%%%%%%%%%%%
\subsection{QGSM results}
      Some information on the charmonium ($c{\bar c}$) 
      and bottomonium ($b{\bar b}$) Regge trajectories
      can be found from the data 
      on the charmed and beauty baryon production in $pp$ 
      collisions at high energies. 
      For example, 
%Fig.~\ref{Fig_3} 
      Fig.3 illustrates the sensitivity of the inclusive spectrum 
      $d\sigma/dx$ of the produced charmed baryons $\Lambda_c$ 
      to different values of the Regge intercept $\alpha_\psi(0)$. 
       Experiment R608 
\cite{Chauvat:1987kb}, after  measuring the decay $\Lambda_c\rightarrow\Lambda^0 + 3\pi$,
       obtained $(2.84 \pm 0.50 \pm 0.72)~\mu$b for the cross section of
%       $pp\rightarrow\Lambda_c X$. 
       $pp\rightarrow\Lambda^0 + 3\pi+ X$
       at $|x|> 0.5$ and $\sqrt{s}=62$ GeV. 
       The branching ratio of this decay is
%       $\Lambda_c\rightarrow\Lambda^0 + 3\pi$ is 
       $(2.8\pm 0.7\pm 1.1)\%$, therefore the cross section 
       of the $\Lambda_c$ production is 
       $\sigma(|x|>0.5)= (101\pm 18\pm 26)~\mu$b. 
       Theoretical expectation for this cross section is 87.3 $\mu$b  
       ($\sigma(|x|>0.5)=87.3~\mu$b)
       with $\alpha_\Psi(0)=0$ and 30.5 $\mu$b with $\alpha_\Psi(0)=-2.18$. 
%       $\sigma(|x|>0.5)=30.5 $\mu$b with $\alpha_\Psi(0)=-2.18$. 
%
       On the other hand, experiment R422 
\cite{Bari:1991in} measured the process 
        $pp\rightarrow e^- \Lambda_c X$ at $|x|>0.35$. 
        With a large uncertainty 
%	(depending on the assumption made in \cite{Bari:1991in}) %, Table 6 in 
	the cross section for the process was obtained to be 
	from ($26\pm 12$) $\mu$b to ($225\pm 9$) $\mu$b \cite{Bari:1991in}. 
%		Therefore, 
        It seems that the open circles (R608 experiment) in Fig.3
%$~\ref{Fig_3}$ 
	better correspond to our calculations (solid line)
	at $\alpha_\Psi(0)=0$.

	Unfortunately, there are no available data 
	for the reaction $pp\rightarrow\Lambda_b X$, some 
        predictions for this kind of reactions are presented in 
\cite{Bednyakov:2010yz}, where it is shown that all the observables are very sensitive 
        to the value of the intercept $\alpha_\Upsilon(0)$ 
	of the $\Upsilon(b{\bar b})$ Regge trajectory. 
	The upper limit of our results is reached at  $\alpha_\Upsilon(0)=0$,
	when this Regge trajectory is the nonlinear $t$-function. 
%~\ref{Fig_5} (left).
\begin{figure}[h!]
%%\begin{center}
%\begin{tabular}{cc}
\hspace{1cm}%
%\mbox
{\epsfig{file=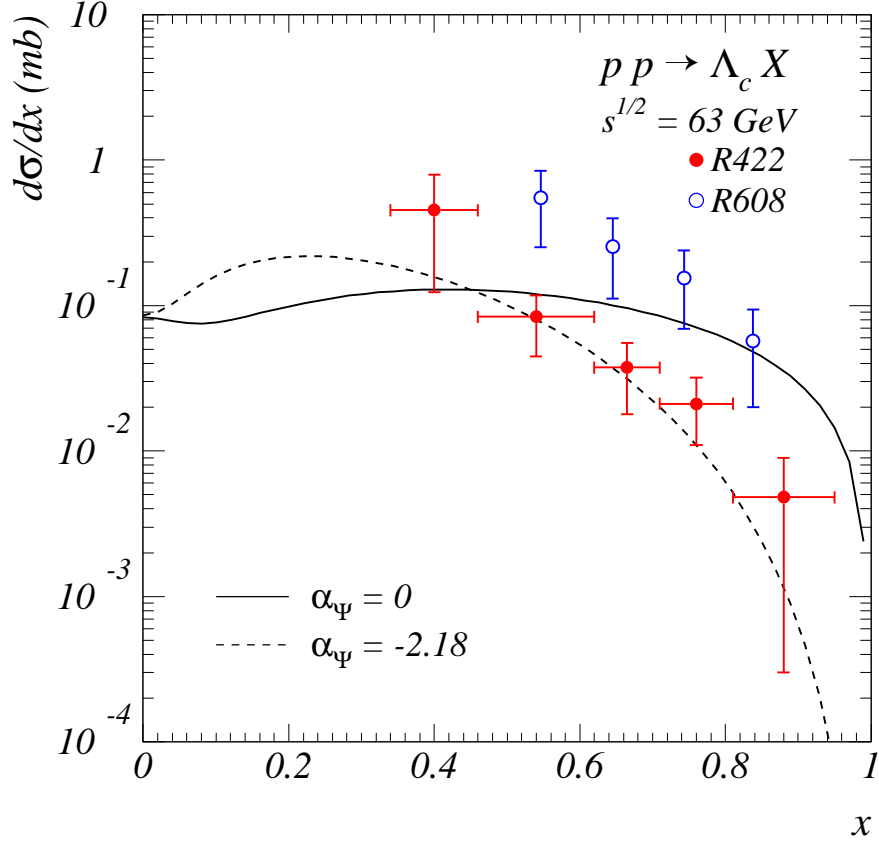,width=0.8\linewidth}}
%&\mbox{\epsfig{file=dCSdx_10TeV.eps,width=0.39\linewidth}}
%\mbox{\epsfig{file=dCSdPt2_10TeV.eps,width=0.43\linewidth}}
%\end{tabular}
%\end{center}
\caption{The differential cross section $d\sigma/dx$ for the
      inclusive process $pp\rightarrow\Lambda_c X$ at $\sqrt{s}=$62 GeV.
%      (left).
      The solid line corresponds to $\alpha_\psi(0)=0$. 
      The dashed curve corresponds to $\alpha_\psi(0)=-2.18$. 
      The open circles correspond to the R608 experiment \cite{Chauvat:1987kb},
      and the dark circles correspond to the R422 experiment \cite{Bari:1991in}.
%      The differential cross section $d\sigma/dx$ 
%          for the process $pp\rightarrow\Lambda_b X$ 
%	 at $\sqrt{s}=$10 TeV (right).
} 
\label{Fig_5}
\end{figure}    
      In fact, to measure the above-mentioned distributions of the 
      process $pp\rightarrow\Lambda_b X$ one should reliably detect the     
      $\Lambda_b$-hyperon. 
      For this purpose, we believe, the beauty baryon decays 
      $\Lambda_b\rightarrow J/\Psi\Lambda^0\rightarrow\mu^+\mu^- p\pi^-$ 
%      (schematically depicted in
(Fig.~\ref{Fig_4}) 
      and $\Lambda_b\rightarrow J/\Psi\Lambda^0\rightarrow e^+e^- n\pi^0$
      can be used.
\begin{figure}[h!]
\centerline{
 {\epsfig{file=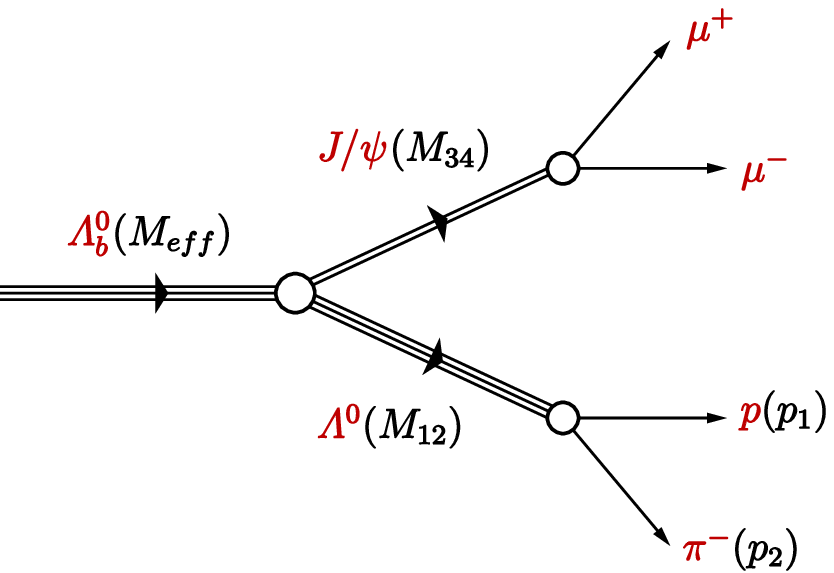,width=0.6\linewidth  }}} 
\caption{The decay $\Lambda_b\rightarrow J/\Psi\Lambda^0\rightarrow \mu^+(e^+) \mu^-(e^-) p(n)\pi^-(\pi^0)$.} 
\label{Fig_4}
\end{figure}
%----------------------------------------------------------------------
         The produced $\Lambda_b$ baryon can undergo the decay 
	 $\Lambda_b\rightarrow J/\Psi \Lambda^0$
	 with the branching ratio 
	 $Br_{\Lambda_b\rightarrow J/\Psi\Lambda^0}=
\Gamma_{\Lambda_b\rightarrow J/\Psi\Lambda^0}/\Gamma_{tot}=(4.7\pm 2.8)\cdot 10^{-4}$. 
          The $J/\Psi$ decays into 
	  $\mu^+\mu^-$ ($Br_{J/\Psi\rightarrow\mu^+\mu_-}=(5.93\pm 0.06)\%$)
	  or into $e^+e^-$ ($Br_{J/\Psi\rightarrow e^+e^-}=(5.94\pm 0.06)\%$), 
	  whereas $\Lambda^0$ can decay into 
	  $p\pi^-$ ($Br_{\Lambda^0\rightarrow p\pi^-}=(63.9\pm 05)\%$), or into $n\pi^0$ 
	  ($Br_{\Lambda^0\rightarrow n\pi^0}=(35.8\pm 0.5)\%$).
 
	  One can experimentally measure the differential cross section of this process 
	  $d\sigma/d\xi_p dt_p dM_{J/\Psi}$, 
	  where $\xi_p=\Delta p/p$ is the energy loss, $t_p=(p_{in}-p_1)^2$ is 
	  the four-momentum transfer, $M_{J/\Psi}$ is the effective mass of the $J/\Psi$-meson.
%
%	Using the hadron detector systems of ATLAS (or CMS and TOTEM) 
        In principle, the ATLAS and CMS detectors
	could detect the decay
$\Lambda^0_b\rightarrow J/\Psi~\Lambda^0\rightarrow \mu^+\mu^-~\pi^0 n(\pi^- p)$
       by recording two muons and one nucleon 
       (neutron in ATLAS - ZDC and proton in TOTEM ) emitted forward. 
       However, the acceptance of the muon detection is 
       $8^0\leq\theta_\mu\leq 172^0$ \cite{Aad:2010ac}-\cite{TOTEM}, 
%\cite{Aad:2010ac,Khachatryan:2010xs,TOTEM}
       and according to our calculations       
        the fraction of expected events in this region is too low. 
       On the other hand, the electromagnetic calorimetry allows one 
       to measure the dielectron pairs $e^+e^-$ with acceptance about
       $1^0\leq\theta_{e(e^+)}\leq 179^0$ \cite{Aad:2010ac}-\cite{TOTEM}.
%\cite{Aad:2010ac,Khachatryan:2010xs,TOTEM}.

\begin{figure}[h!]
%\begin{center}
%\begin{tabular}{cc}
\hspace{0.5cm}%
%\mbox
{\epsfig{file=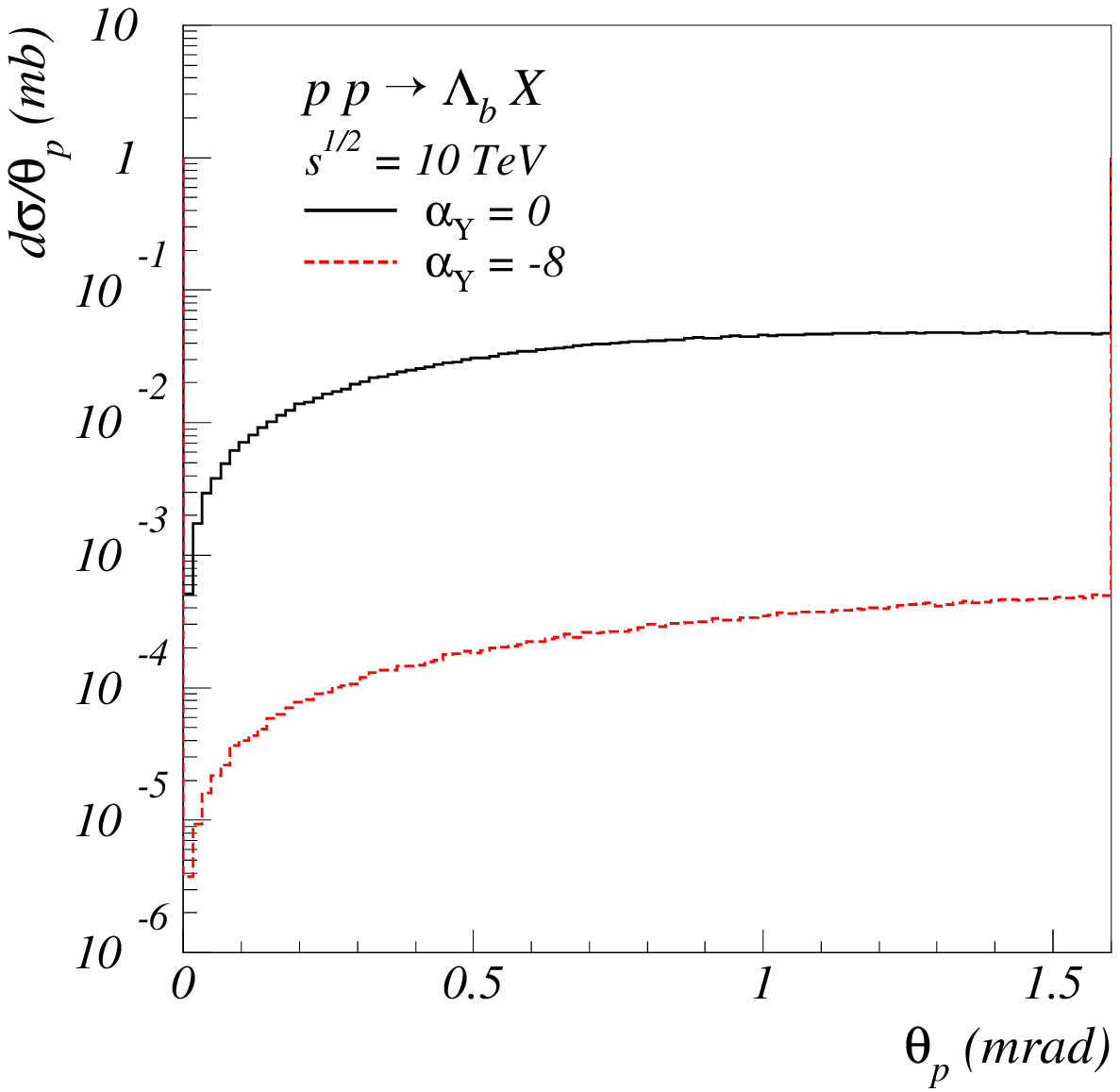,width=0.8\linewidth}}
 \caption{The distributions over $\theta_p$   
          for $pp\rightarrow\Lambda_b X\rightarrow\mu^+\mu^- p\pi^- X$ 
          at $\sqrt{s}=10$ TeV.
          The solid (long dashed) curve corresponds to $\alpha_\Upsilon(0)=0$
          ($\alpha_\Upsilon(0)=-8$).
}
\end{figure}
\begin{figure}[h!]
%\mbox
{\epsfig{file=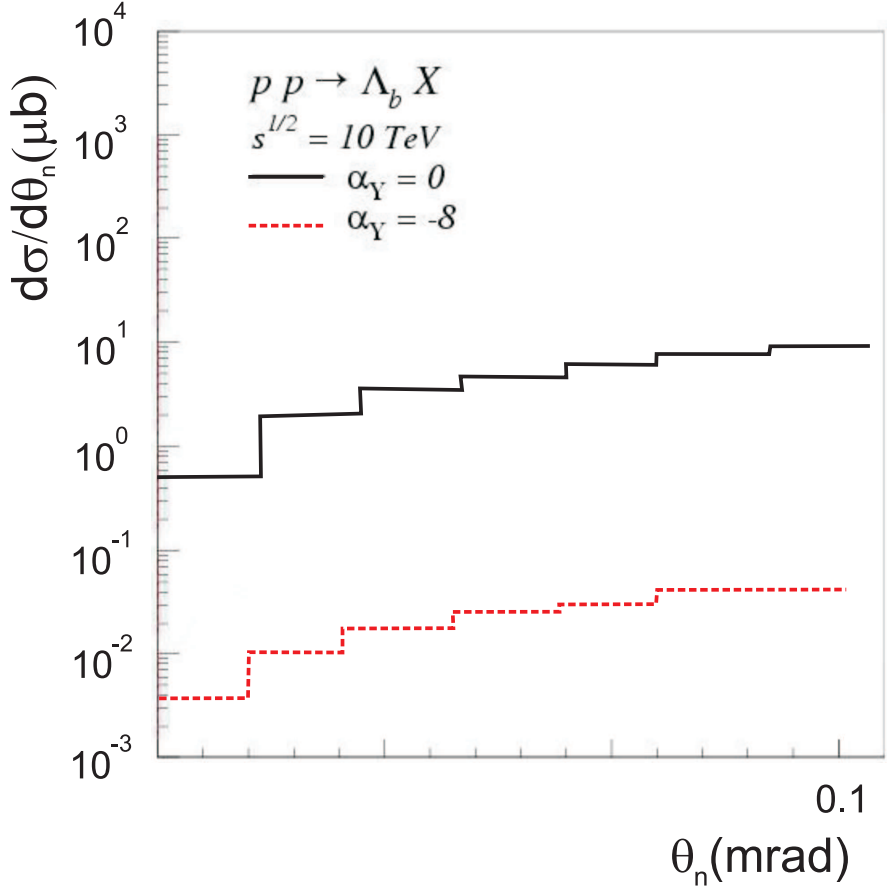,width=0.8\linewidth}}
%\end{tabular}
%\end{center}
 \caption{ The distribution $d\sigma/d\theta_n$ over the neutron scattering angle $\theta_n$  
          for $pp\rightarrow\Lambda_b X\rightarrow e^+e^-\pi^0 n~X$ at $\sqrt{s}=10~\mathrm{TeV}$.
          The solid (long dashed) curve corresponds to $\alpha_\Upsilon(0)=0$
          ($\alpha_\Upsilon(0)=-8$).  
}
\label{Fig_7}
\end{figure}
      In Figs.(5,6)
      the  distributions over  the proton scattering angle $\theta_p$ in the reactions 
      $pp\rightarrow\Lambda_b X\rightarrow J/\Psi\Lambda_0 X\rightarrow \mu^+\mu^- p\pi^- X$
      and over the neutron scattering angle $\theta_n$ in the process
      $pp\rightarrow\Lambda_b X\rightarrow J/\Psi\Lambda_0 X\rightarrow e^+e^-\pi^0 n~X$ 
      are presented at the intercept values $\alpha_\Upsilon(0)=0$ (solid line) and
      $\alpha_\Upsilon(0)=-8$ (dashed line). 
      Figures~5,6 show a large sensitivity of these distributions
      to the intercepts $\alpha_\Upsilon=0$ and $\alpha_\Upsilon=-8$ of the $\Upsilon(b{\bar b})$ Regge 
      trajectory. One can see that the cross section $d\sigma/d\theta_{p,n}$ for the nonlinear  $\Upsilon(b{\bar b})$ 
      Regge trajectory ($\alpha_\Upsilon=0$) is larger than for the linear one ($\alpha_\Upsilon=-8$).  
      As is shown above for the $\Lambda_c$ production in $pp$ collisions, the use of the nonlinear $\psi(c{\bar c})$ 
      Regge trajectory results in the better correspondence of the calculations to the experimental data. Therefore, 
      one can assume that all our results obtained at  $\alpha_\Upsilon=0$ are more preferable than the ones at 
      $\alpha_\Upsilon=-8$ and they can be considered as the upper limit of our QGSM calculations.         
      The reaction 
      $pp\rightarrow\Lambda_b X\rightarrow J/\Psi\Lambda_0 X\rightarrow\mu^+\mu^-(e^+e^-) p\pi^- X$
      can be measured by the TOTEM and CMS, and the neutron produced from the 
      $\Lambda_0$ decay 
      can be be detected by the ATLAS Zero Degree Calorimeter (ZDC).

	 In 
Figs.(7,8) the two-dimensional plots of the energy and scattering angle for the positron (electron)
        produced in the $J/\Psi$ decay and the neutron produced in the $\Lambda^0$ decay are presented. 
        These plots correspond to $\alpha_\Upsilon=0$.    	

\begin{figure}[h!]
%\begin{center}
%\begin{tabular}{cc}
%\mbox
{\epsfig{file=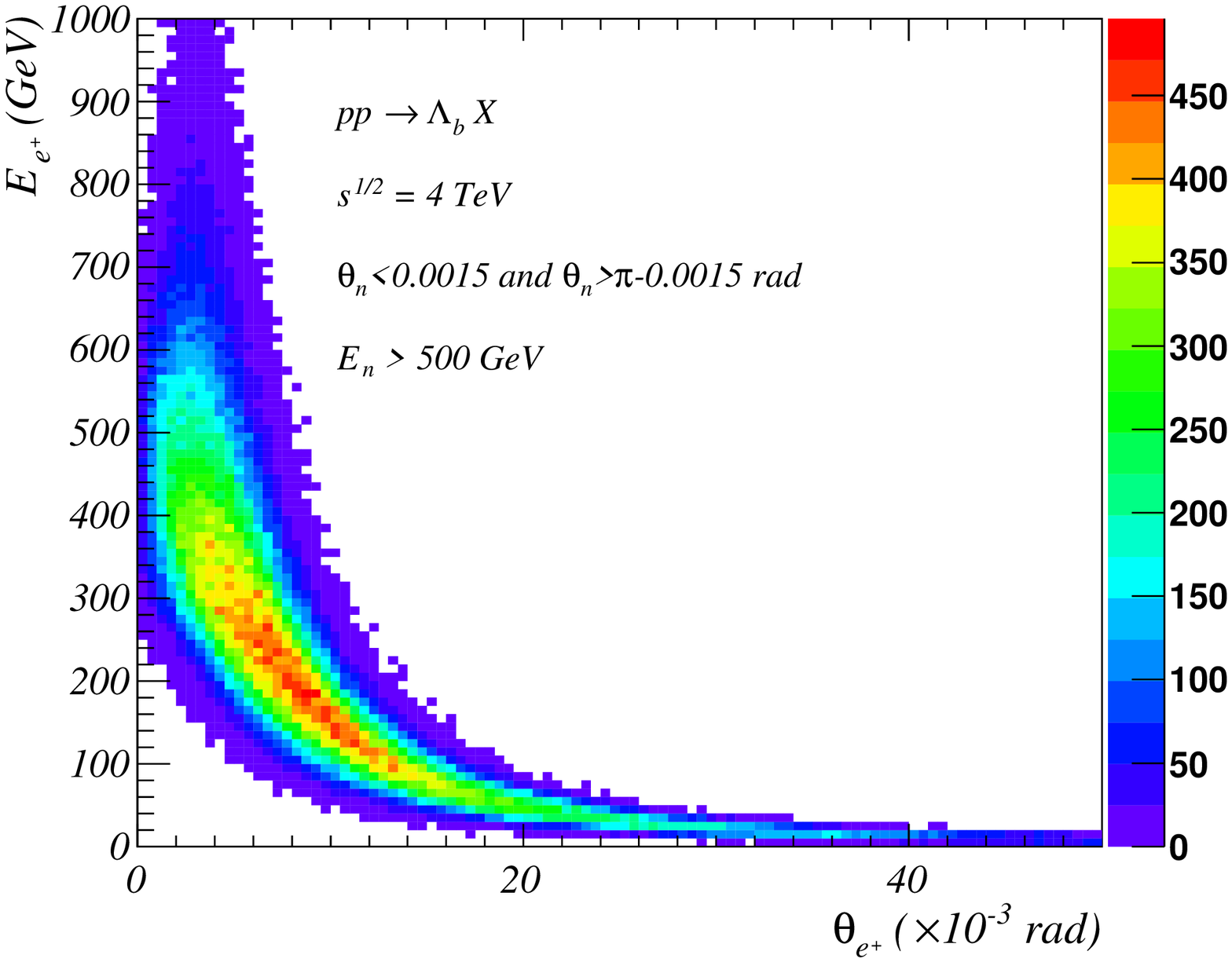,width=0.9\linewidth}}
%\mbox
 \caption{The distributions over $\theta_{e^+}$ and $E_{e^+}$ in the 
         inclusive process $pp\rightarrow\Lambda_b X\rightarrow J/\Psi\Lambda^0 X
	 \rightarrow e^+e^- n\pi^0 X$ at $\sqrt{s}=4~$TeV. 
	 The rate of these events is about 4.6 percent (13.8 nb). }
\label{Fig_8}
\end{figure}
\begin{figure}[h!]
{\epsfig{file=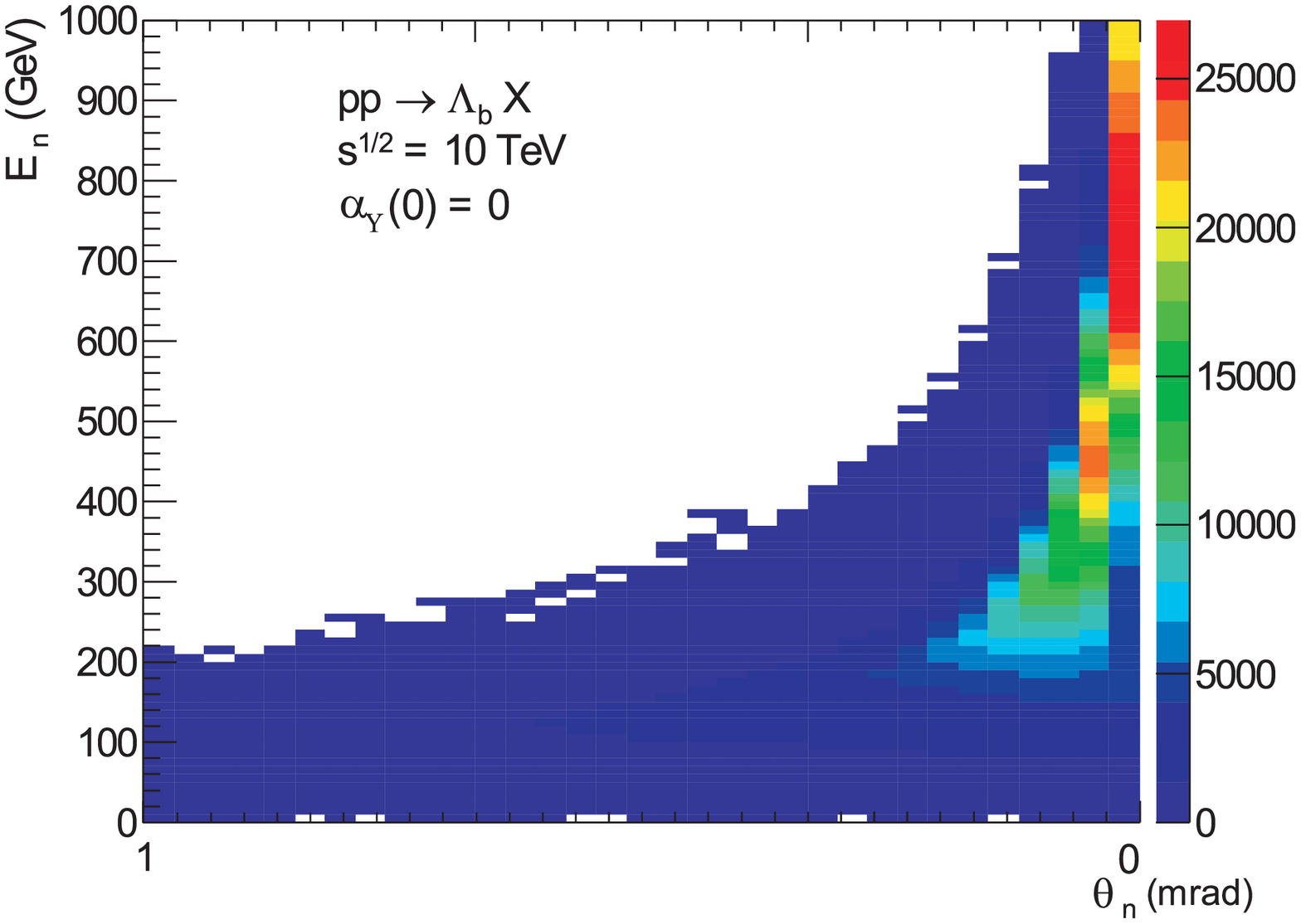,width=0.9\linewidth}}
%\end{tabular}
%\end{center}
 \caption{
 The neutron energy distribution as a function of $\theta_n$ for the
         process $pp\rightarrow\Lambda_b X\rightarrow e^+e^-\pi^0 n~X$ at $\sqrt{s}=10~\mathrm{TeV}$.       
}
\label{Fig_9}
\end{figure}
        One can see from 
Figs.~(7,8)
        that positrons (electrons) are concentrated 
        at the scattering angles 4 mrad$<\theta_{e*+}<$16 mrad, 
	whereas the neutrons are emitted mainly at $\theta_n\leq~$0.1 mrad
	which could be measured with 
	ATLAS using the ZDC (n) or with the TOTEM $\&$ CMS ($p$). 
        The ATLAS experiment 
	is also able to detect $e^+e^-$ by  
	the electromagnetic calorimeter in the interval 
	$1^0\leq\theta_{e(e^+)}\leq 179^0$ 
\cite{Aad:2010ac}.
         The ratio of the events presented in Fig.~(7) to the 
         total yield (let us call it the rate) is about  4.6 percent (13.8 nb).
    The detailed analysis and theoretical predictions for the $\Lambda_b$ production in the forward
    $pp$ collisions without the inclusion of the possible intrinsic bottom in the proton were presented
    in \cite{Bednyakov:2010yz,Artemenkov:2010wj}. It was shown \cite{Artemenkov:2010wj} that the rate 
    of the events in the reaction 
    $pp\rightarrow\Lambda_b\rightarrow J/\Psi~\Lambda^0\rightarrow e^+e^-~\pi^- p$ at $\sqrt{s}=7~\mathrm{TeV}$
    when 500 GeV$\le E_p\le$ 4 TeV and the proton scattering angle 
    3 mrad~$\le \theta_p\leq~$ 10 mrad was about 1.33 $\%$ (40 nb). In principle, this kinematics corresponds to 
    that of the TOTEM and CMS facilities at CERN. And the rate  in the reaction 
    $pp\rightarrow\Lambda_b\rightarrow J/\Psi~\Lambda^0\rightarrow e^+e^-~\pi^0 n$ at the same initial energy     
    when $E_n\leq$ 3 TeV and the neutron scattering angle $\theta_n\leq~$ 0.1 mrad is about 0.0207 (60. pb).
    This kinematics corresponds to the ATLAS and ZDC at CERN.
 
    Therefore, we see that the cross section of the $\Lambda_b$-hyperon produced in $pp$ collisions and 
    decayed into $\pi^- p$ or $\pi^0 n$ calculated within the QGSM without the intrinsic bottom in the proton
    can be from 60 pb to 40 nb at the different kinematics of the LHC experiments at $\sqrt{s}=7~\mathrm{TeV}$. 
\subsection{Intrinsic charm and beauty contribution}
\smallskip
      Let us discuss the opportunity to find some information on the 
      distributions of intrinsic charm and beauty in the proton 
      from the analysis of the forward production of
      heavy flavour baryons in $pp$ collisions at the LHC. 
      First, we note that all sea quark distributions in the proton 
      calculated within the QGSM give their contributions
      only to the multi-Pomeron graphs at $n\geq 2$  
(Fig.~\ref{Fig.2}~(right)). 
      According to Eq.~(\ref{Eq_2}),  %~(2).
      they do not contribute to the one-Pomeron graph  
(Fig.~\ref{Fig.2}~(left)). 
       The one-Pomeron graph results in the main contribution at large 
       values of $x$ because $\sigma_n$ decreases very fast when $n$ increases 
\cite{TerMartirosyan:1973yn}. 
       Therefore %, as our calculations showed recently 
       the sea charm and beauty quark distributions %, in fact, 
       result in very small contributions to the inclusive 
       spectra of the charmed and beauty hadrons 
       at not large $p_t$ 
       and different values of $x$
\cite{Lykasov:2008md,Lykasov:2009zd,Bednyakov:2010yz}. 
      For the forward heavy flavour hadron production
      this contribution is smaller and we neglect it. 

      These {\em sea}\/ charm and beauty quark distributions
      greatly differ from %are not the same as 
      the distributions of the {\em intrinsic}\/ charm and bottom quarks 
      in the proton, 
      which, according to the assumption of   
\cite{ Brodsky:1980pb,Pumplin:2005yf}, 
      should behave like the {\em valence}\/ 
      quark distributions.
      Therefore, if one wants to estimate the contribution of   
      the intrinsic charm and bottom quarks within the QGSM scheme,
      one should include their distributions 
      in the calculation of the one-Pomeron graph ($n=1$). 
      The procedure can increase the rates of events for both reactions 
      $pp\rightarrow\Lambda_b X\rightarrow e^+e^- p\pi^- X$
      and $pp\rightarrow\Lambda_b X\rightarrow e^+e^- n\pi^0 X$
      when the final proton or neutron is emitted in the forward direction 
      because the one-Pomeron graph 
      (Fig.~\ref{Fig.2}~(left)) 
      makes the major contribution to the $\Lambda_b$ spectrum  
      in this kinematics.

    Assuming the existence of the intrinsic $b{\bar b}$-pair in the proton,
      as a pair of the {\it valence} quark-antiquark ,  
      with some non-zero probability $w_{b{\bar b}}$, one can 
      estimate enhancement in the forward $\Lambda_b$ $pp$-production at the LHC.       
     The expected enhancement will be about the ratio of the differential cross 
     sections   
     $\frac{d\sigma}{dx}(pp\to n \,X)/ \frac{d\sigma}{dx}(pp\to\Lambda_b X)$   
     at very large $x$ multiplied by $w_{b{\bar b}}$,
      Although it is suppressed in comparison with the intrinsic
      charm probability $w_{c{\bar c}}$ 
      by a factor $m^2_c/m^2_b\simeq 0.1$
\cite{Polyakov:1998rb}, it can not be neglected  nevertheless. 

     Calculating these spectra within the QGSM \cite{Kaidalov:1985jg,Bednyakov:2010yz,Lykasov:2009zd}
     at large $x$ and assuming $w_{b{\bar b}}\sim 0.3\%$ \cite{Nadolsky:2008zw,Polyakov:1998rb} 
     one can get that   
     the yield of $\Lambda_b$ produced hadronically in the forward direction 
     can increase by a factor 3-5 times %by a factor a few times (3-5), when 
     due to the intrinsic bottom quark contribution. %is included. 
     It means that the cross section of the forward production of $\Lambda_b$ in $pp$ 
     collisions at LHC energies, which decays into $e^+e^- \pi^- p$ or $e^+e^- \pi^0 n$,
     can reach a few hundred nb for TOTEM and CMS and few hundred pb 
     for ATLAS. 
     Our estimations show that the yield of the forward charmed $\Lambda_c^+$-hyperon  
     production 
     can be increased 
     by a factor of 10 %or more 
     due to the intrinsic charm quarks.
     Therefore, the reaction 
$pp\rightarrow \Lambda_c^+ X\rightarrow\Lambda^0\pi^+ X\rightarrow n\pi^0\pi^+ X$
     can also be measured at the LHC 
     when the neutron is emitted in the forward direction. 
     The neutron can be measured by
     the ZDC and $\pi^+$-meson can be detected by the hadron calorimeter.
\begin{figure}[h!]
%\begin{center}
\begin{tabular}{cc}
\hspace{0.5cm}%
{\epsfig{file=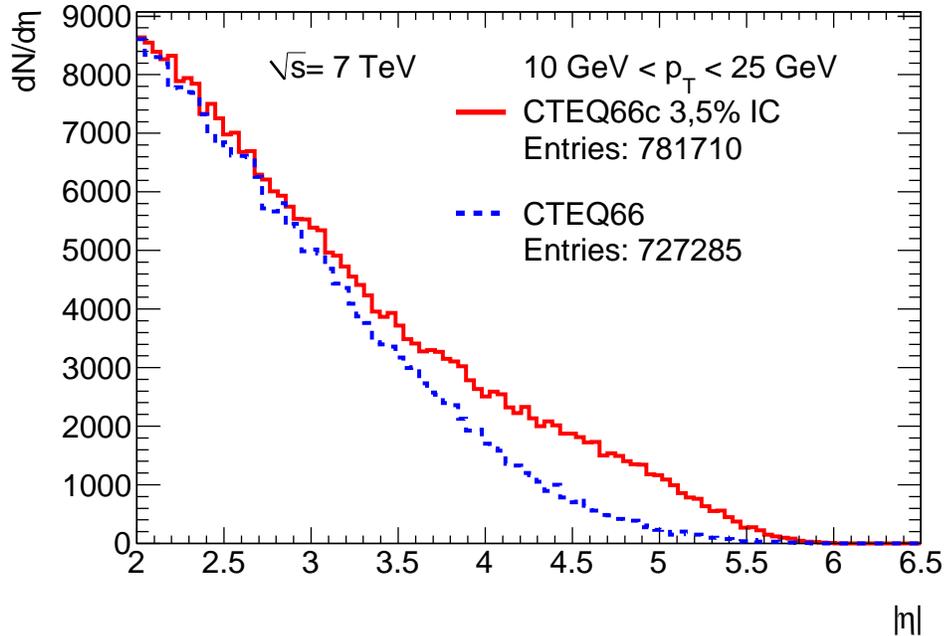,width=0.88\linewidth}}
%\mbox{\epsfig{file=pp_D0D0X.eps,width=0.45\linewidth}}
\end{tabular}
%\end{center}
 \caption{The $D+{\bar D}_0$ distributions over the pseudo-rapidity $\eta$ in 
          $pp\rightarrow (D_0+{\bar D}_0) X$  
          at $\sqrt{s}=7$ TeV and $10\leq p_t\leq 25$ GeV$/$c. }
\label{Fig_7}
\end{figure}
\begin{figure}[h!]
%\rotatebox{270}
 {\epsfig{file=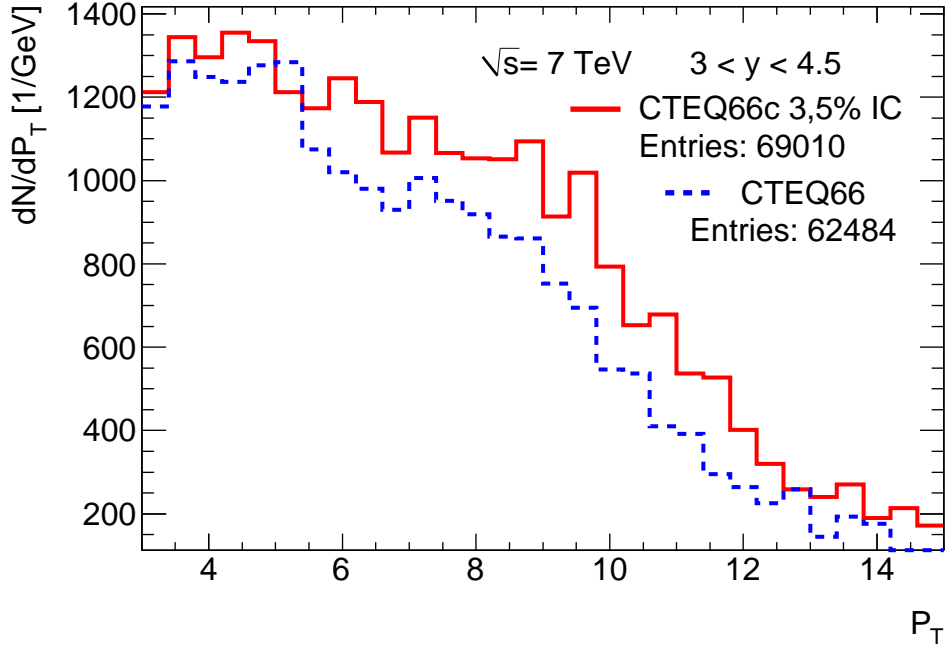,width=0.9\linewidth  }}
\caption{The double $D^0$ distributions over the pseudo-rapidity $\eta$ in 
          $pp\rightarrow (D_0+{\bar D}_0) X$ and at $p_t\geq 25$ GeV$/$c.} 
\label{Fig_8}
\end{figure}
We presented the qualitative estimations for the contributions of the intrinsic beauty and charm
to the forward $\Lambda_b$ and $\Lambda_c$ production at LHC. These spectra were calculated
within the nonperturbative QGSM in which the PDF do not include the intrinsic charm or beauty 
contributions. It is not so easy to take into account the intrinsic charm (IC) contribution at 
the PDF used in the QGSM \cite{Kaidalov:1982xg,Kaidalov:1985jg}.

However, there are the PDF used in the perturbative QCD calculations  which include the IC contribution
in the proton \cite{Pumplin:2005yf}-\cite{Nadolsky:2008zw}.
The probability distribution for the 5-quark state ($uudc{\bar c}$) in the light-cone description of the proton
was first calculated in \cite{ Brodsky:1980pb}. It has the following form 
\cite{Pumplin:2007wg}
%{Brodsky:1980pb,Brodsky:1981} 
\begin{eqnarray}
\frac{dP}{dx}=f_c(x)=f_{\bar c}(x)={\cal N}x^2\times
%\nonumber \\
\left\{(1-x)(1+10x+x^2)+6x(1+x)ln(x)\right\}~,
\label{def:fcPumpl}
\end{eqnarray} 
where ${\cal N}$ is the normalization constant. One can see from Eq.(\ref{def:fcPumpl}) that
the IC distribution has some enhancement at large $x$ and vanishes at $x=1$. 
As is shown in \cite{Nadolsky:2008zw} this enhancement starts at $x~>~0.2$ 
that can result in similar enhancement
in the inclusive spectra of the open charm at large rapidities $y$ or pseudorapidities $\eta$
and transverse momenta $p_t$. We calculated 
the IC contribution to the inclusive spectra of the $D$-mesons produced in $pp$ collisions 
at $\sqrt{s}=$7 TeV within the perturbative QCD. In Fig.9 the inclusive spectrum of single $D^0$-mesons is
presented as a function of the pseudorapidity $\eta$ at 10 GeV$/$c$<p_t<$25 GeV$/$c. 
Calculating these spectra within PYTHIA8 we used the PDF both for the CTEQ66 without the IC 
(the dashed blue distributions in Fig.9)
and the CTEQ66c including the IC with the probability about 3.5 $\%$ (the solid red distributions 
in Fig.(9))  at $Q^2=m_c^2=$1.69 GeV$^2$ \cite{Nadolsky:2008zw}.
Figure 9 shows that some enhancement due to the IC can be visible at large $\eta$. Its amount increases
when $p_t$ grows and, for example, the inclusion of the IC increases the spectrum by a factor of 2 at 
$\eta=$4.5.
Similar  effect was predicted in 
\cite{Kniehl:2012ti}.
In Fig.10 the inclusive spectrum of the $D^0$-meson produced in the process $pp\rightarrow D^0D^0 X$
at the rapidity interval 3$<y<$4.5 and  $\sqrt{s}=$7 TeV is presented as a function of $p_t$, when two $D^0$ 
are produced. 
The enhancement of the spectrum (the excess of the solid histogram 
in comparison to the dashed one) at 7$<p_t<$10 GeV$/$c is not more than 30 $\%$. The predictions
presented in Figs.(9,10) can be verified by the LHCb experiment at CERN because their facility is able to measure
inclusive spectra of $D$-mesons at $\eta\leq$ 4.5.
\section{Conclusion}
          We analyzed production of charmed and beauty baryons 
	  in proton-proton collisions at high energies 
	  within the soft QCD quark-gluon string model. 
	  This approach described 
	  rather satisfactorily the charmed baryon production in $pp$ collisions 
\cite{Lykasov:2009zd,Bednyakov:2010yz} and allowed us to apply the 
	  QGSM to beauty baryon production. 
	  We focus mainly on the analysis of the forward $\Lambda_b$ 
	  production in $pp$ collisions at LHC energies.   
	  We present the predictions for the reaction 
	  $pp\rightarrow\Lambda_b X\rightarrow e^+e^- n\pi^0 X$,
	  which can be studied in %reliable at 
	  the ATLAS experiment using the ZDC, %\cite{ATLAS1,ZDC} 
	  and for the process
	  $pp\rightarrow\Lambda_b X\rightarrow e^+e^- p\pi^- X$, 
	  which could be reliably studed at the TOTEM and CMS
          within their common proposal for Diffractive and 
          Forward Physics at the LHC \cite{Khachatryan:2010xs,TOTEM}. 
%\cite{TOTEM,Deile}.
	  We show that beauty $\Lambda_b$-hyperons 
	  can in principle 
	  be detected in LHC experiments via 
	  registration %(in forward direction) 
	  of their  
	  decay products $\Lambda_b \rightarrow e^+e^- p\pi^-$
	  and $\Lambda_b \rightarrow e^+e^- n\pi^0$.

	  We would like to stress that any data on forward 
	  production in the process %$d\sigma/dx$ and $d\sigma/dP_t^2$ 
	  $pp\rightarrow\Lambda_b X$ could be important 
	  for determination of $\alpha_\Upsilon(0)$
	  of the $\Upsilon(b{\bar b})$ Regge trajectory.
	  Our predictions with the nonlinear Regge trajectories of $c{\bar c}$
	  and $b{\bar b}$ mesons can be considered as 
	  the upper limit of the QGSM calculations 
	  without the intrinsic charm and bottom in the proton.

	  The inclusion of the intrinsic bottom or/and charm in the proton 
	  can increase the yield of the relevant 
	  heavy flavour baryons by a factor of 3 to 10. 
	  In particular, we considered a possibility of measuring the reaction 
	  $pp\rightarrow \Lambda_c^+ X\rightarrow\Lambda^0\pi^+ X\rightarrow n\pi^0\pi^+ X$ 
	  using the ATLAS (ZDC). 
	  This measurement can provide
	  information on the intrinsic charm in the proton, 
	  the probability of which is estimated to be a factor of 10
	  higher than the one for the intrinsic bottom in the proton. 
	  Finally, it is worth noticing that any reliable non-observation 
	  of this enhancement 
	  in the experiments at the LHC 
	  can severely constrain the intrinsic heavy quark hypothesis. 

          Our calculations of the charmed meson production in $pp$ collisions within the
          MC generator PYTHIA8 and the PDF including the intrinsic charm showed the following.
          We found that the contribution of the intrinsic charm in the proton could be
          studied in the production of $D$-mesons in $pp$ collisions at the LHC. The IC contribution
          for the single $D^0$-meson production can be sizable, it is about 100 $\%$ at large 
          rapidities 3 $\leq y\leq$ 4.5 and large transverse momenta 10 $\leq p_t\leq$ 25 GeV$/$c. 
          For the double $D^0$ production this contribution is not larger than 30 $\%$ at 
          $p_t\geq$ 5 GeV$/$c and  3 $\leq y\leq$ 4.5. These IC contributions for the single and 
          double $D$-meson production were obtained with the probability of the intrinsic charm 
          taking to be $w_{c{\bar c}}=$3.5 $\%$ \cite{Nadolsky:2008zw}, and they will decrease 
          by a factor of 3 when $w_{c{\bar c}}\simeq$ 1 $\%$. Therefore, this value can be verified 
          experimentally at LHCb. 

          The presented predictions could stimulate measurement of the single and double
          D-meson production in $pp$ collisions at the CERN LHCb experiment  in the kinematic region 
          mentioned above to observe a possible signal for the intrinsic charm. The intrinsic beauty
          in the proton is suppressed by a factor of 10, therefore its signal in the inclusive spectra
          of $B$-mesons will probably be very weak.
              
\section{Acknowledgments}
We are very grateful to V.V.~Lyubushkin and D.A.~Artemenkov 
for their help with the MC calculations.
We also thank  M. Deile   
for extremely useful suggestions related to the possible experimental check of our 
predictions at the LHC.
We are also grateful to I.Belyaev, V.Gligorov, D.Denegri, K. Eggert, H.Jung, B.Kniehl,
S.Lami, T.Lomtadze, A.Likhoded, F.Palla, P.Spradlin,
O.V.Teryaev, M. Poghosyan and S.White for very useful discussions. 
This work was supported in part by the Russian Foundation for Basic Research 
grant N: 11-02-01538-a.

%To acknowledge funding bodies etc., a special section may be placed
%before the bibliography: \verb?\section*{Acknowledgements}?.
% ****************************************************************************
% BIBLIOGRAPHY AREA
% ****************************************************************************
% please do not change the following line
%\begin{footnotesize}
%\bibliographystyle{eplbib} 
%\bibliography{ic}
%\end{footnotesize}

\end{document}